\documentclass[times, authoryear, 11pt]{elsarticle}

\usepackage{setspace}
\usepackage{amssymb}

\usepackage{geometry}
\usepackage{wrapfig}
\usepackage{caption}
\usepackage{amsmath}
\usepackage{hyperref}
\usepackage{cleveref}
\usepackage{algorithm}
\usepackage[noend]{algpseudocode}
\usepackage{setspace}
\algrenewcommand\algorithmiccomment[1]{\hfill #1}
\newcommand{\R}{\mathbb{R}}
\newcommand{\E}{\mathbb{E}}
\newcommand{\ub}[1]{^{(#1)}}
\newcommand{\mat}[1]{\mathrm{#1}}
\newcommand{\id}{\mathbb{I}}
\newcommand{\T}{\mathrm{T}}
\newcommand{\cov}{\textrm{Cov}}
\newcommand{\gauss}{\mathbf{N}}

\DeclareMathAlphabet{\mathcal}{OMS}{cmsy}{m}{n}

\newcommand*\return{\State \textbf{return} }

\hypersetup{
  colorlinks=true,
}

\begin{document}

\begin{frontmatter}



\title{Ensemble Kalman Inversion for General Likelihoods}


\author{Samuel Duffield \corref{cor1}\fnref{label1}}


\author{Sumeetpal S. Singh}

\address{Department of Engineering, University of Cambridge, CB2 1PZ, UK}

\fntext[label1]{Corresponding author: s@mduffield.com
}


\begin{abstract}
In this letter we generalise Ensemble Kalman inversion techniques to general Bayesian models where previously they were restricted to additive Gaussian likelihoods - all in the difficult setting where the likelihood can be sampled from, but its density not necessarily evaluated.
\end{abstract}




\end{frontmatter}


\section{Introduction}\label{sec:intro}

\textit{Approximate Bayesian computation} (ABC) refers to a collection of Monte Carlo methods for inference in Bayesian models
\begin{equation}\label{bayes_model}
    x \sim p(x), \qquad y \sim p(y\mid x),
\end{equation}
where $x$ is an unknown parameter to be inferred, $y$ is some data that is known, $p(x)$ is a prior distribution and $p(y \mid x)$ is a likelihood function describing the data generating process. With ABC, we have the further restriction that the likelihood function cannot be evaluated pointwise and therefore the likelihood is said to be \textit{intractable} - this rules out classical techniques such as Markov chain Monte Carlo (MCMC) or importance sampling. Instead, ABC techniques rely on the ability to simulate synthetic data from the likelihood $y\ub{i} \sim p(y \mid x\ub{i})$. Here the superscript indicates that $(x\ub{i}, y\ub{i})$ are the $i$th pair from a collection of $N$ parameter-simulated data pairs $\{(x\ub{i}, y\ub{i})\}_{i=1}^N$.
\par
The majority of ABC techniques have focused on adapting classical techniques to target an extended distribution that can be sampled from without evaluating the likelihood.
Simulated data $\{y\ub{i}\}_{i=1}^N$ can then be compared to the true observed data $y$ and subsequently only those parameter values $\{x\ub{i}\}_{i=1}^N$ with generated data sufficiently \textit{close} to the true data are retained (where the measure of closeness is typically defined by a combination of summary statistics, distance function and tolerance level).
Popular ABC variants are based on rejection sampling \citep{Pritchard1999}, MCMC \citep{Marjoram2003} or sequential Monte Carlo (SMC) \citep{Jasra2011}, whilst there have also been approaches to incorporate machine learning techniques \citep{Lueckmann2019} and alternative posteriors \citep{Matsubara2021}. For a thorough review of ABC techniques, see \cite{Beaumont2019}.
ABC methods are asymptotically biased, however this bias is typically quantifiable and controllable (at the expense of further computational cost). 
\par
This setting of intractable likelihoods is challenging. As yet, ABC methods have been restricted to relatively low-dimensional Bayesian models and are renowned for requiring many likelihood simulations and therefore long run-times. The development of efficient and scalable ABC methods remains a prominent research goal.
\par
In contrast, ensemble Kalman techniques \citep{Evensen1994, Iglesias2013} are a class of Monte Carlo algorithms that have become very popular for inference in high-dimensional models with the restriction of additive Gaussian likelihoods
\begin{equation}\label{nonlin_bip}
    x \sim p(x), \qquad y  \sim \gauss (y \mid \mathcal{H}(x), \mat{R}),
\end{equation}
where $\mathcal{H}$ a deterministic forward operator and $\mat{R}$ a noise covariance matrix which are both assumed known. As noted in \cite{Nott2012}, ensemble Kalman methods make no use of likelihood evaluations and therefore they are themselves ABC techniques in the specific setting of additive Gaussian noise \eqref{nonlin_bip}.
\par
In this letter, we remove this restriction and \textbf{apply ensemble Kalman techniques to general likelihoods} \eqref{bayes_model}. Thus developing an extremely general algorithm for inference in Bayesian models with intractable likelihoods that can perform well even with few likelihood simulations and is scalable to higher dimensions compared to conventional ABC techniques. 
The rest of the letter is structured as follows. \Cref{sec:enk} provides a brief account of existing ensemble Kalman techniques. We then introduce our generalised ensemble Kalman method in \Cref{sec:genk} before examining its performance numerically for both optimisation and uncertainty quantification in \Cref{sec:sims}. \Cref{sec:conc} concludes and discusses future work.

\section{Ensemble Kalman Approach}\label{sec:enk}

Ensemble Kalman techniques were first introduced \citep{Evensen1994} for generating online Monte Carlo approximations for state-space models.

For simplicity, let us consider the offline setting of \eqref{nonlin_bip} where where the likelihood takes the form $p(y \mid x) = \gauss (y \mid \mathcal{H}(x), \mat{R})$ with $x\in \R^{d_x}$ and $y\in \R^{d_y}$. Then a single update step of the ensemble Kalman filter as described in \cite{Houtekamer2001} takes the form
\begin{align}\label{eki_single}
\begin{split}
    x_1\ub{i} &= x_{0}\ub{i} + \mat{C}^{x\mat{H}}_{0} \left( \mat{C}^{\mat{HH}}_{0} + \mat{R} \right)^{-1}(y - \mathcal{H}(x_{0}\ub{i}) - \eta\ub{i}), \\
    \eta\ub{i} &\sim \gauss (\eta \mid 0, \mat{R}),
\end{split}
\end{align}
with prior samples $x_0\ub{i} \sim p(x)$, empirical covariance matrices
\begin{align}\label{h_covs}
\begin{split}
    \mat{C}^{x\mat{H}}_0 &= \frac1{N-1} \sum_{i=1}^N(x_0\ub{i} - \bar{x}_0)(\mathcal{H}(x_0\ub{i}) - \bar{\mathcal{H}}_0)^{\T}, \\
    \mat{C}^{\mat{HH}}_0 &= \frac1{N-1} \sum_{i=1}^N (\mathcal{H}(x_0\ub{i}) - \bar{\mathcal{H}}_0) (\mathcal{H}(x_0\ub{i}) - \bar{\mathcal{H}}_0)^{\T},
\end{split}
\end{align}
and empirical means $\bar{x}_0 = \frac1N \sum_{i=1}^N x_0\ub{i}$ and $\bar{\mathcal{H}}_0 = \frac1N \sum_{i=1}^N \mathcal{H}(x_0\ub{i})$.
\par
This ensemble Kalman update is asymptotically unbiased in the special case of Gaussian prior $p(x) = \gauss (x \mid m, \mat{Q})$ and linear Gaussian likelihood $p(y \mid x) = \gauss (y \mid \mat{H} x, \mat{R})$ which gives
\begin{align}
    p(x,y)
    &=
    \gauss
    \left(
    \begin{pmatrix}
    x \\ y
    \end{pmatrix}
    \mid
    \begin{pmatrix}
    m \\ \mat{H}m
    \end{pmatrix}
    ,
    \begin{pmatrix}
    \mat{Q} & \mat{QH}^{\T} \\
    \mat{HQ} & \mat{HQH}^{\T} + \mat{R}
    \end{pmatrix}
    \right).\label{lg}
\end{align}
This model has an analytically tractable posterior distribution $p(x \mid y) = \gauss(x \mid m^y, \mat{Q}^y)$ where
\begin{align*}
    m^y &= m + \mat{QH}^{\T} \left( \mat{HQH}^{\T} + \mat{R} \right)^{-1}(y - \mat{H}m),\\
    \mat{Q}^y &= \mat{Q} - \mat{QH}^{\T} \left( \mat{HQH}^{\T} + \mat{R} \right)^{-1} \mat{HQ}.
\end{align*}
Outside of this special linear Gaussian case, the ensemble Kalman update \eqref{eki_single} is known to be asymptotically biased. There is however, a vast amount of empirical evidence demonstrating stability in a variety of complex non-linear state-space models with a very small number of particles, e.g. \cite{Houtekamer2005}, \cite{Roth2017}. In our view, this summarises the ensemble Kalman paradigm:
\begin{align*}
\begin{minipage}{.8\textwidth}
    \centering
    \bf
    Asymptotically unbiased in the linear Gaussian case, otherwise biased but empirically stable.
\end{minipage}
\end{align*}
Where the term \textit{empirically stable} represents the ability of the ensemble $\{x^{(i)}\}_{i=1}^N$ to cover the true underlying value of the state without degenerating to a single particle. Naturally, the acceptance of a bias induced by this paradigm could be extreme in some problems, however as mentioned, there is still very much a practical desire for these numerically efficient but biased methods in this difficult setting of intractable likelihoods.
\par
For non-Gaussian priors and non-linear Gaussian likelihoods \eqref{bayes_model}, the distribution of particles from a single step ensemble Kalman update may be quite different from the true posterior. As noted in \cite{Iglesias2013}, this can be mitigated by iterating the ensemble Kalman update in \eqref{eki_single}. The idea was refined in \cite{Iglesias2015} to be more in line with the tempered likelihood approach of SMC samplers \citep{DelMoral2006, Jasra2011}. This iterative ensemble Kalman approach is termed \textit{ensemble Kalman inversion} (EKI) and a single iterate takes the form
\begin{align}\label{eki}
\begin{split}
    x_\ell\ub{i} &= x_{\ell-1}\ub{i} + \mat{C}^{x\mat{H}}_{\ell-1} \left( \mat{C}^{\mat{HH}}_{\ell-1} + h_\ell^{-1}\mat{R} \right)^{-1}(y - \mathcal{H}(x_{\ell-1}\ub{i}) - \eta\ub{i}_{\ell}), \\ \eta\ub{i}_{\ell} &\sim \gauss (\eta_\ell \mid 0, h_\ell^{-1}\mat{R}),
\end{split}
\end{align}
where $\mat{C}^{x\mat{H}}_{\ell-1}$ and $\mat{C}^{\mat{HH}}_{\ell-1}$ are as in \eqref{h_covs} applied to the particles $\{ x_{\ell-1}\ub{i} \}_{i=1}^N$, the subscript indicates the iteration of the algorithm, i.e. $\ell=0, \dots, L$ where $L$ is the total number of iterations.
The parameter $h_\ell^{-1}$ can be considered a stepsize parameter or equally $h_\ell$ as an incremental inverse temperature parameter. In the fully linear Gaussian case
\eqref{lg}, the particles $\{x_\ell\ub{i}\}_{i=1}^N$ at each step are asymptotically unbiased for a tempered version of the posterior
\begin{align}\label{lg:tempered_post}
    p_\ell(x) &\propto p(x)p(y\mid x)^{\lambda_\ell} \propto \gauss(x \mid m_\ell, \mat{Q}_\ell), \nonumber\\
    m_\ell &= m + \mat{QH}^{\T} \left( \mat{HQH}^{\T} + \lambda_\ell^{-1}\mat{R} \right)^{-1}(y - \mat{H}m), \\
    \mat{Q}_\ell &= \mat{Q} - \mat{QH}^{\T} \left( \mat{HQH}^{\T} + \lambda_\ell^{-1}\mat{R} \right)^{-1} \mat{HQ}, \nonumber
\end{align}
where $\lambda_\ell = \sum_{r=1}^\ell h_r$ is the inverse temperature. The tempered posterior can also be defined iteratively
\begin{align}\label{lg:tempered_post_rec}
\begin{split}
    m_\ell &= m_{\ell-1} + \mat{Q}_{\ell-1}\mat{H}^{\T} \left( \mat{H}\mat{Q}_{\ell-1}\mat{H}^{\T} + h_\ell^{-1}\mat{R} \right)^{-1}(y - \mat{H}m_{\ell-1}), \\
    \mat{Q}_\ell &= \mat{Q}_{\ell-1} - \mat{Q}_{\ell-1}\mat{H}^{\T} \left( \mat{H}\mat{Q}_{\ell-1}\mat{H}^{\T} + h_\ell^{-1}\mat{R} \right)^{-1} \mat{H}\mat{Q}_{\ell-1}.
\end{split}
\end{align}
Thus, iterating for $L$ steps with stepsizes such that $\lambda_L = \sum_{r=1}^L h_r = 1$ obtains particles $\{x_L\ub{i}\}_{i=1}^N$ that are asymptotically unbiased for the true posterior in the linear Gaussian case.
\par
Iterating until $\lambda_L = 1$ is only one possible method of termination, another approach is to adopt an optimisation style stopping criterion, for example \cite{Iglesias2015} suggest terminating when the average of the forward operations $\bar{\mathcal{H}}_L$ is suitably close the true data, e.g. when
$(y - \bar{\mathcal{H}}_L)^\T \mat{R}^{-1} (y - \bar{\mathcal{H}}_L) < \tau$ for some stopping parameter $\tau$.

\section{Generalised Ensemble Kalman Inversion}\label{sec:genk}

In this section, we generalise ensemble Kalman inversion to the case where data is generated according to any likelihood $p(y \mid x)$ rather than the setting described in \Cref{sec:enk} which is restricted to likelihoods of the form $\gauss (y \mid \mathcal{H}(x), \mat{R})$. The resulting algorithm only requires samples from the prior $p(x)$ and the ability to simulate from the likelihood $p(y \mid x)$ for a given parameter $x$, as in approximate Bayesian computation. The final particle approximation is asymptotically biased for the true posterior with the exception of the fully linear Gaussian case \eqref{lg} and therefore remains faithful to the ensemble Kalman paradigm.
 
\subsection{General Likelihoods}
In this more general case, we can no longer form the empirical covariance matrices in \eqref{h_covs} as we do not have access to the deterministic $\mathcal{H}$. Instead we can form
\begin{subequations}\label{eki:y_covs}
\begin{align}
    \mat{C}^{xx}_\ell &= \frac1{N-1} \sum_{i=1}^N(x_\ell\ub{i} - \bar{x}_\ell)(x_\ell\ub{i} - \bar{x}_\ell)^{\T}, \\
    \mat{C}^{xy}_\ell &= \frac1{N-1} \sum_{i=1}^N(x_\ell\ub{i} - \bar{x}_\ell)(y_\ell\ub{i} - \bar{y}_\ell)^{\T}, \\
    \mat{C}^{yy}_\ell &= \frac1{N-1} \sum_{i=1}^N (y_\ell\ub{i} - \bar{y}_\ell)(y_\ell\ub{i} - \bar{y}_\ell)^{\T},
\end{align}
\end{subequations}
 and $\mat{C}^{yx}_\ell = \mat{C}^{xy\, \T}_\ell$. Here we have simulated data $y_\ell\ub{i} \sim p( \cdot \mid x_\ell\ub{i})$ and means $\bar{x}_\ell = \frac1N \sum_{i=1}^N x_\ell\ub{i}$, $\bar{y}_\ell = \frac1N \sum_{i=1}^N y_\ell\ub{i}$.
 \par
 In the fully linear Gaussian case \eqref{lg} or rather the tempered version (\ref{lg:tempered_post}, \ref{lg:tempered_post_rec}) we have $\mat{C}^{xy}_\ell \to \mat{HQ}_\ell$ and $\mat{C}^{yy}_\ell \to \mat{HQ}_\ell \mat{H}^\T + \mat{R}$.
 We now note that we can also calculate
 \begin{align}
     \mat{C}^{y|x}_\ell = \mat{C}^{yy}_\ell - \mat{C}^{yx}_\ell \mat{C}^{xx \, -1}_\ell \mat{C}^{xy}_\ell, \label{emp_r}
 \end{align}
and observe that in the linear Gaussian case $\mat{C}^{y|x}_\ell \to (\mat{H} \mat{Q}_{\ell-1}\mat{H}^\T + \mat{R}) - 
    (\mat{H} \mat{Q}_{\ell-1}) \mat{Q}_{\ell-1}^{-1} (\mat{Q}_{\ell-1} \mat{H}^\T) 
    = \mat{R}.$
Thus we can use the particles to approximate the covariance of the likelihood empirically.
We therefore adjust the tempered ensemble Kalman iteration \eqref{eki} to the following
\begin{align}
    x_\ell\ub{i} &= x_{\ell-1}\ub{i} + \mat{C}^{xy}_{\ell-1} \left( \mat{C}^{yy}_{\ell-1} + (h_\ell^{-1}-1)\mat{C}^{y|x}_{\ell-1} \right)^{-1}\left(y - y_{\ell-1}\ub{i} - \eta\ub{i}_{\ell}\right), \nonumber\\
    y_{\ell-1}\ub{i} &\sim p(y_{\ell-1} \mid x_{\ell-1}\ub{i}), \label{new_eki}\\
    \eta\ub{i}_{\ell} &\sim \gauss (\eta \mid 0, (h_\ell^{-1} - 1)\mat{C}^{y|x}_{\ell-1}). \nonumber
\end{align}
In the linear Gaussian case and large sample limit, the above ensemble Kalman move becomes
\begin{align*}
    x_\ell\ub{i} &= x_{\ell-1}\ub{i} + \mat{Q}_{\ell-1}\mat{H}^\T \left( (\mat{H} \mat{Q}_{\ell-1}\mat{H}^\T + \mat{R}) + (h_\ell^{-1}-1)\mat{R} \right)^{-1}(y - y_{\ell-1}\ub{i} - \eta\ub{i}_{\ell}),\\
    y_{\ell-1}\ub{i} &\sim \gauss (y_{\ell-1} \mid \mat{H}x_{\ell-1}\ub{i}, \mat{R}),\\
    \eta\ub{i}_{\ell} &\sim \gauss (\eta_\ell \mid 0, (h_\ell^{-1} - 1)\mat{R}).
\end{align*}
Where the $(h_\ell^{-1} - 1)$ scaling has been chosen to ensure that the statistics $\E[x_\ell\ub{i}]$ and $\cov[x_\ell\ub{i}]$ match those in \eqref{lg:tempered_post_rec}.
Therefore the ensemble Kalman update in \eqref{new_eki} is asymptotically unbiased in the linear Gaussian case, and consistent with the ensemble Kalman paradigm. The full procedure is listed in Algorithm~\ref{alg:new_eki}.

\begin{algorithm}
\caption{Ensemble Kalman Inversion for General Likelihoods}
\label{alg:new_eki}
\begin{algorithmic}[1]
\State Given a sequence of inverse temperatures $\{\lambda_\ell\}_{\ell = 0}^L$  and stopping criterion  - which may both be adaptive, see Sections \ref{eki:stepsize_sel} and \ref{eki:stopping}.
\State Sample from prior
\begin{equation*}
    x_0^{(i)} \sim p(x) \tag*{$i=1,\dots,N$}
\end{equation*}
\For{$\ell=1,\dots,L$}
\State Simulate observations
\begin{equation*}
    y_{\ell-1}^{(i)} \sim p(y \mid x_{\ell-1}^{(i)}) \tag*{$i=1,\dots,N$}
\end{equation*}
\State Form sample covariances $\mat{C}^{xx}_{\ell-1}, \mat{C}_{\ell-1}^{xy}, \mat{C}^{yy}_{\ell-1}$ \eqref{eki:y_covs} and $\mat{C}^{y|x}_{\ell - 1}$ \eqref{emp_r}.
\State Set stepsize $h_\ell = \lambda_\ell - \lambda_{\ell-1}$
    \State Generate observation perturbations
    \begin{equation*}
        \eta_{\ell}^{(i)} \sim \gauss\left(\eta \mid 0, \left( h_\ell^{-1}-1\right)\mat{C}^{y|x}_{\ell-1} \right) \tag*{$i=1,\dots,N$}
    \end{equation*}
    \State Move particles
    \begin{equation*}
        x_\ell\ub{i} = x_{\ell-1}\ub{i} + \mat{C}^{xy}_{\ell-1} \left( \mat{C}^{yy}_{\ell-1} + (h_\ell^{-1}-1)\mat{C}^{y|x}_{\ell-1} \right)^{-1}\left(y - y_{\ell-1}\ub{i} - \eta\ub{i}_{\ell}\right)
        \tag*{$i=1,\dots,N$}
    \end{equation*}
\EndFor
\return $\{ x_L^{(i)} \}_{i=1}^N$
\end{algorithmic}
\end{algorithm}

\subsection{Stepsize Selection}\label{eki:stepsize_sel}
The motivation for iterative ensemble Kalman inversion is that for difficult non-Gaussian problems, moving directly from prior to posterior in one step is an extremely difficult task. By taking many smaller steps the particles can explore the state-space and have a better chance of settling in regions of high posterior probability.
There is, therefore, a trade-off to be made - more steps means greater exploration but also more likelihood simulations and a longer runtime.
\par
The stepsizes $\{h_\ell\}_{\ell=1}^L$ equivalently define an inverse temperature schedule $0 = \lambda_0 < \lambda_1 < \dots <\lambda_L$ where $h_\ell = \lambda_\ell - \lambda_{\ell -1} >0$ for $\ell = 1, \dots, L$.
\par
For SMC samplers \citep{DelMoral2006}, it is common for the next inverse temperature to be selected adaptively \citep{Jasra2011} such that the effective sample size (of the sequential importance weights $\{w\ub{i}_\ell\}_{i=1}^N$) decreases by a fixed amount at each iteration. That is, select $\lambda_\ell$ such that $\text{ESS}( \{w\ub{i}_\ell\}_{i=1}^N ) = (\sum_{i=1}^N w^{(i) \, 2}_\ell)^{-1} \approx \rho N$, where the normalised weights are a function of $\lambda_\ell$ and $\rho \in (0,1)$ is a tuning parameter that controls the size of the steps. The root for $\lambda_\ell$ can be found using a numerical bisection algorithm in $(\lambda_{\ell-1}, \lambda_L]$ and requires no additional likelihood evaluations. This way the algorithm smoothly transitions from prior to posterior whereas simply using a constant stepsize might induce a degenerate series of intermediate distributions unless an excessively small stepsize is used. This idea was ported to ensemble Kalman inversion in \cite{Iglesias2018} where the following psuedo-weights are used 
$
\hat{w}^{(i)}_\ell \propto \exp \left(-\frac12 (\lambda_\ell - \lambda_{\ell -1}) \left(y - \mathcal{H}( x_{\ell-1}^{(i)})\right)^\T \mat{R}^{-1}\left(y -\mathcal{H}( x_{\ell-1}^{(i)})\right)\right),$
as ensemble Kalman inversion does not compute importance weights inherently. Here we directly adapt these pseudo-weights to the general likelihood case
\begin{equation} \label{eki_pseudo_w}
    \hat{w}^{(i)}_\ell \propto \exp \left(-\frac12 (\lambda_\ell - \lambda_{\ell -1}) \left(y - y_{\ell-1}^{(i)}\right)^\T \mat{C}^{y|x \, -1}_{\ell-1}\left(y -y_{\ell-1}^{(i)}\right)\right).
\end{equation}
\par
As noted in \cite{Iglesias2018}, the lack of resampling means the user can be more aggressive with the choice of $\rho$, in our experiments we set $\rho = \frac12$.



\subsection{Stopping Criteria}\label{eki:stopping}
We consider two stopping criteria.

\begin{itemize}
    \item \textbf{Sampling} - terminate when $\lambda_L = 1$. This stopping criterion mimics that of tempered likelihood approaches for tractable likelihoods and is applied to ensemble Kalman inversion in \cite{Iglesias2018}. This approach is asymptotically unbiased for the true posterior in the linear Gaussian case and aims to quantify uncertainty around parameters.
    
    
    \item \textbf{Optimisation} - terminate when the marginal variance (of the particles) in each dimension becomes sufficiently small, i.e. when $[\mat{C}^{xx}_\ell]_{kk} < \upsilon [\mat{C}^{xx}_0]_{kk}, \forall k \in \{1,\dots d_x\}$ for a predefined $\upsilon \in [0,1]$ (which we set to $10^{-2}$). Here the notation $[A]_{ij}$ indexes the $i,j$ coordinate in the matrix $A$. Under this approach the algorithm is iterated until the particles form a consensus approaching a single point estimate, which in the linear Gaussian case will (asymptotically) be the maximum likelihood estimator.
\end{itemize}

\section{Numerical Experiments}\label{sec:sims}

We now examine the performance of both ensemble Kalman inversion for sampling and optimisation versus approximate Bayesian computation methods for two static Bayesian inference problems with intractable likelihoods.
\par
We consider two ABC techniques, the first of which being that of ABC-SMC \citep{DelMoral2012}. In ABC-SMC, a series of importance weighted Monte Carlo approximations are generated iteratively to target

\begin{equation}\label{abc_post}
    \nu_{\kappa_\ell}(x, y^\prime) \propto p(x) p(y^\prime \mid x) \id[||y - y^\prime|| < \kappa_\ell],
\end{equation}

for a series of decreasing parameters $\kappa_0 > \kappa_1 > \dots > \kappa_L$. Here $y$ is the true data, $y^\prime$ is simulated data, whilst $\mathbb{I}[\cdot]$ is the indicator function and $||\cdot ||$ is the Euclidean norm. As in \cite{DelMoral2006}, particles are rejuvenated using a random-walk Metropolis-Hastings kernel, with covariance adapted to $2.38^2 d_x^{-1}\mat{C}^{xx}_{\ell-1}$ a scaled version of the empirical covariance of the previous particles and threshold parameters $\kappa_\ell$ adaptively chosen such that ESS $\approx0.9N$. Particles are resampled when the ESS drops below $0.5N$ and the algorithm terminates when the acceptance rate of the Metropolis-Hastings kernel first falls below 1.5\%.
\par
For our second ABC technique, we run the random-walk Metropolis-Hastings kernel directly as an ABC-MCMC algorithm. As described in \cite{Vihola2020} we use a Robbins-Monro schedule to adaptively tune the stepsize, preconditioner combination to $2.38^2 d_x^{-1} \hat{\mat{\Sigma}}$ where $\hat{\mat{\Sigma}}$ is the empirical covariance of the historical chain and adapt the threshold parameter $\kappa$ from \eqref{abc_post} such that 10\% of samples are accepted. 
\par
As mentioned, for EKI we determine the stepsize $h_\ell$ (equivalently the next inverse temperature $\lambda_\ell$) adaptively such that the ESS of the pseudo-weights \eqref{eki_pseudo_w} is approximately $0.5N$. We examine both the sampling and optimisation stopping criteria discussed in \Cref{eki:stopping}. All simulations are ran with mocat \cite{Duffield2021}, which includes open source implementations for all the discussed algorithms.

\subsection{$g$-and-$k$ distribution}
Popular as a benchmark for ABC algorithms, the g-and-k distribution family is defined by the quantile function
\begin{equation} \label{gk_quantile}
    F^{-1}(u) = A + B \left(1 + c\frac{1 - \exp(-g z(u))}{1 + \exp(-g z(u))} \right) (1 + z(u)^2)^k z(u),
\end{equation}
where $z(\cdot)$ is the quantile function of a standard normal distribution. The constant $c$ is typically set to 0.8 and considered known. We set the remaining parameters $x = (A, B, g, k)$ with \textit{true} values $(3, 1, 2, 1/2)$ but consider them unknown (to be inferred) each with prior $\mathcal{U}(\cdot \mid 0, 10)$. The g-and-k likelihood is defined implicitly by the quantile function \eqref{gk_quantile}. This likelihood function cannot be obtained analytically and thus we cannot easily utilise traditional posterior inference methods such as MCMC or importance sampling - although their ABC counterparts are still possible as the likelihood can be easily simulated for given parameters by simply evaluating the quantile function \eqref{gk_quantile} at a sampled standard uniform random variable.
\par
We assume we have data of 1000 i.i.d. samples from \eqref{gk_quantile} which we summarise into 100 evenly spaced order statistics for both EKI and ABC. It is common practice for Monte Carlo algorithms to apply a transformation converting constrained variables to ones in $\mathbb{R}$, this alleviates the possibility of parameters being moved outside of the bounded domain, here we unconstrain all parameters using the transformation $z(\cdot / 10)$ for both ABC and EKI (other transformations are possible although the choice is not typically considered a significant factor in performance).

\begin{figure}
    \centering
    \begin{minipage}{.48\textwidth}
    \captionsetup{width=.8\linewidth}
    \centering
    \includegraphics[width=\textwidth]{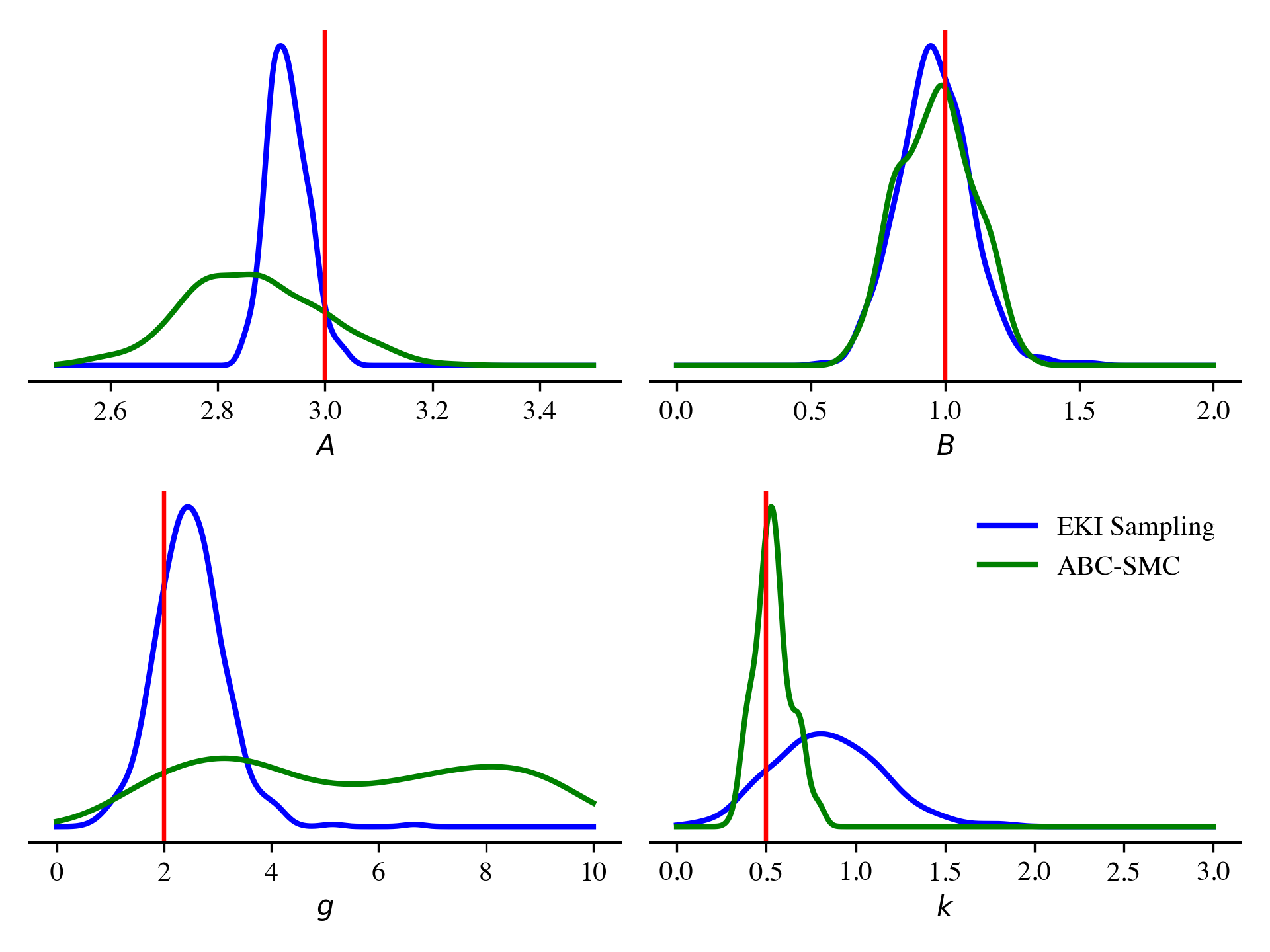}
    \caption{$g$-and-$k$ marginal posteriors for EKI sampling and ABC-SMC, $N=500$, truth in red.}
    \label{fig:gk_densities}
    \end{minipage}%
    \begin{minipage}{.48\textwidth}
    \captionsetup{width=.8\linewidth}
    \centering
    \includegraphics[width=\textwidth]{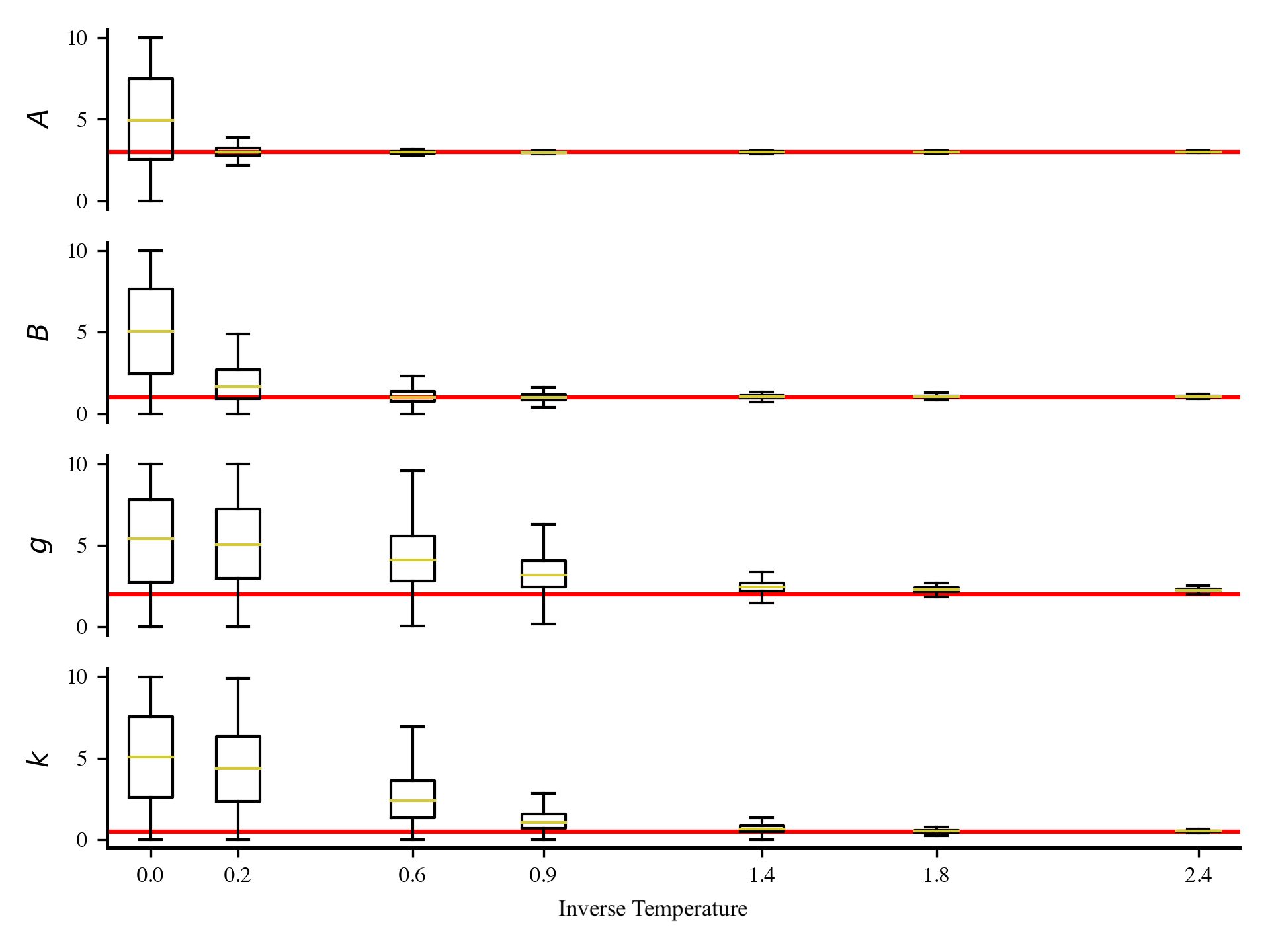}
    \caption{$g$-and-$k$ convergence of EKI for optimisation, $N=500$, truth in red. Boxes display range and quartiles of ensemble.}
    \label{fig:gk_eki_bps}
    \end{minipage}
\end{figure}



In \Cref{fig:gk_densities}, we compare the marginal distributions produced by EKI for sampling and those from ABC-SMC. We observe that EKI has centred on the vicinity of the truth for all 4 parameters whereas ABC-SMC has failed to concentrate in the $g$ variable in particular. We note that the two posteriors differ quite significantly - an indication of the high levels of non-linearity in the $g$-and-$k$ likelihood - yet the EKI posterior remains informative.

We then examine the second EKI stopping criterion, that of optimisation. The EKI optimisation procedure, \Cref{eki:stopping}, iterates beyond $\lambda_\ell = 1$ until a consensus is reached on a single point approximation for the maximum likelihood estimator. \Cref{fig:gk_eki_bps} displays 7 equally spaced iterations as the inverse temperature (which is adaptively selected at each iteration based on the distance of between simulated and true data \eqref{eki_pseudo_w}) increases above $\lambda_\ell=1$ before terminating when all standard deviations are suitably small, note the increasing stepsizes as the ensemble converges to a point estimate. The particles converge quickly and very closely to the true underlying parameter values.

\begin{wrapfigure}{l}{0.5\textwidth}
    \centering
    \includegraphics[width=0.48\textwidth]{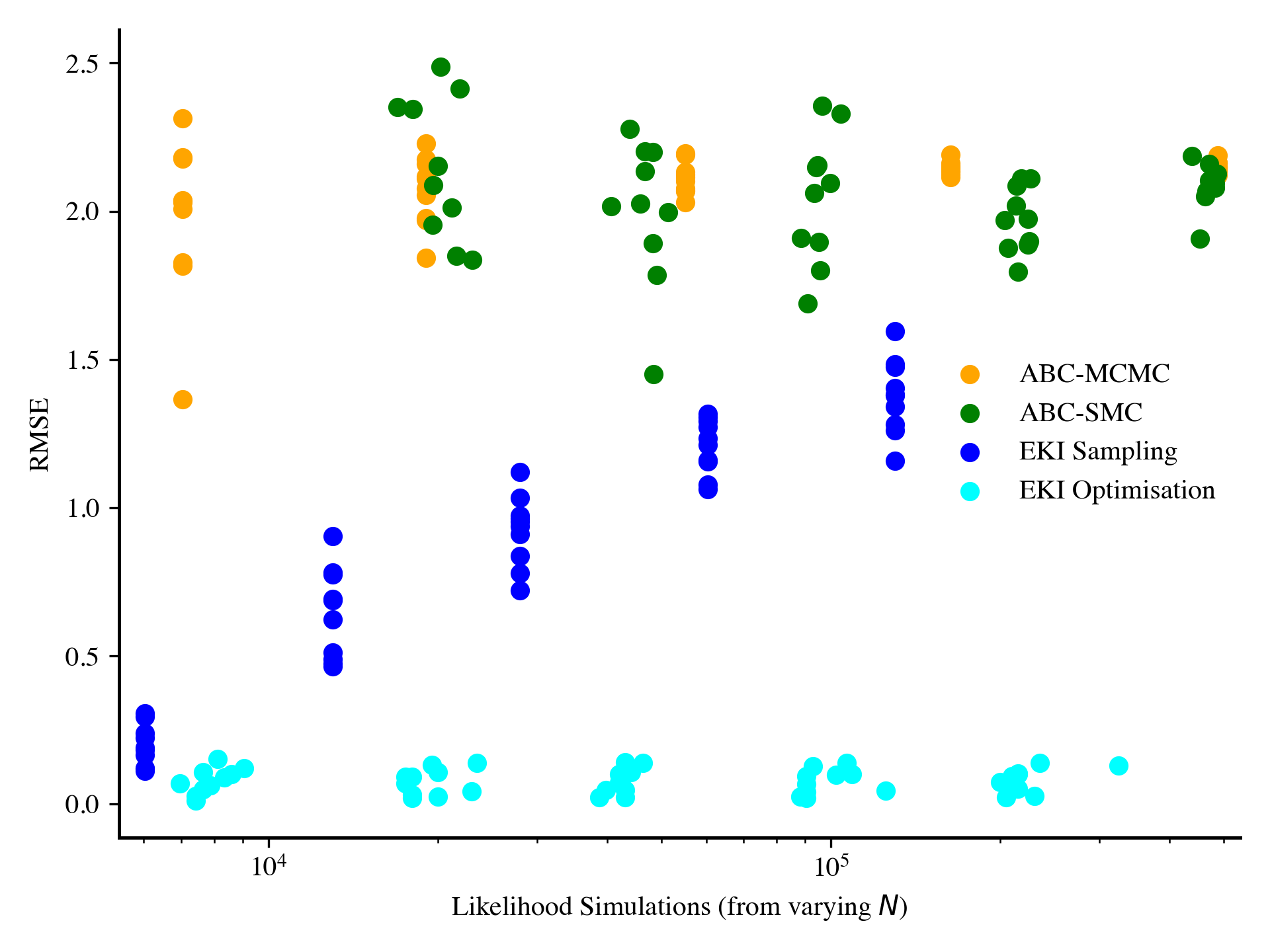}
    \caption{Root mean squared error versus number of likelihood simulations induced by varying $N$ (from 200 to 5000), on $g$-and-$k$ distribution. Repeated over 10 randomly generated sets of observations.}
    \label{fig:gk_rmse}
\end{wrapfigure}

Finally, we vary the sample size $N$ between 200 and 5000 then repeat the experiment 10 times.
We plot the root mean squared error against the number of likelihood simulations utilised by each algorithm as $N$ changes in \Cref{fig:gk_rmse}. Note that the adaptive stopping of the EKI and ABC-SMC techniques means that the number of iterations and therefore likelihood simulations can vary across seeds. We observe that both EKI stopping criteria are consistently outperforming both ABC-MCMC and ABC-SMC for the same computational cost. Curiously, we observe that the performance of EKI for sampling deteriorates as $N$ increases although this is not the case for the optimisation approach. We posit that this might be due to the smaller sample sizes under estimating the noise covariance (a common occurrence in ensemble Kalman techniques \cite{Tong2016}) and therefore moving the particles further. 

\subsection{Stochastic Lorenz 96}

\begin{figure}
    \centering
    \begin{minipage}{.48\textwidth}
    \captionsetup{width=.8\linewidth}
    \centering
    \includegraphics[width=0.8\textwidth]{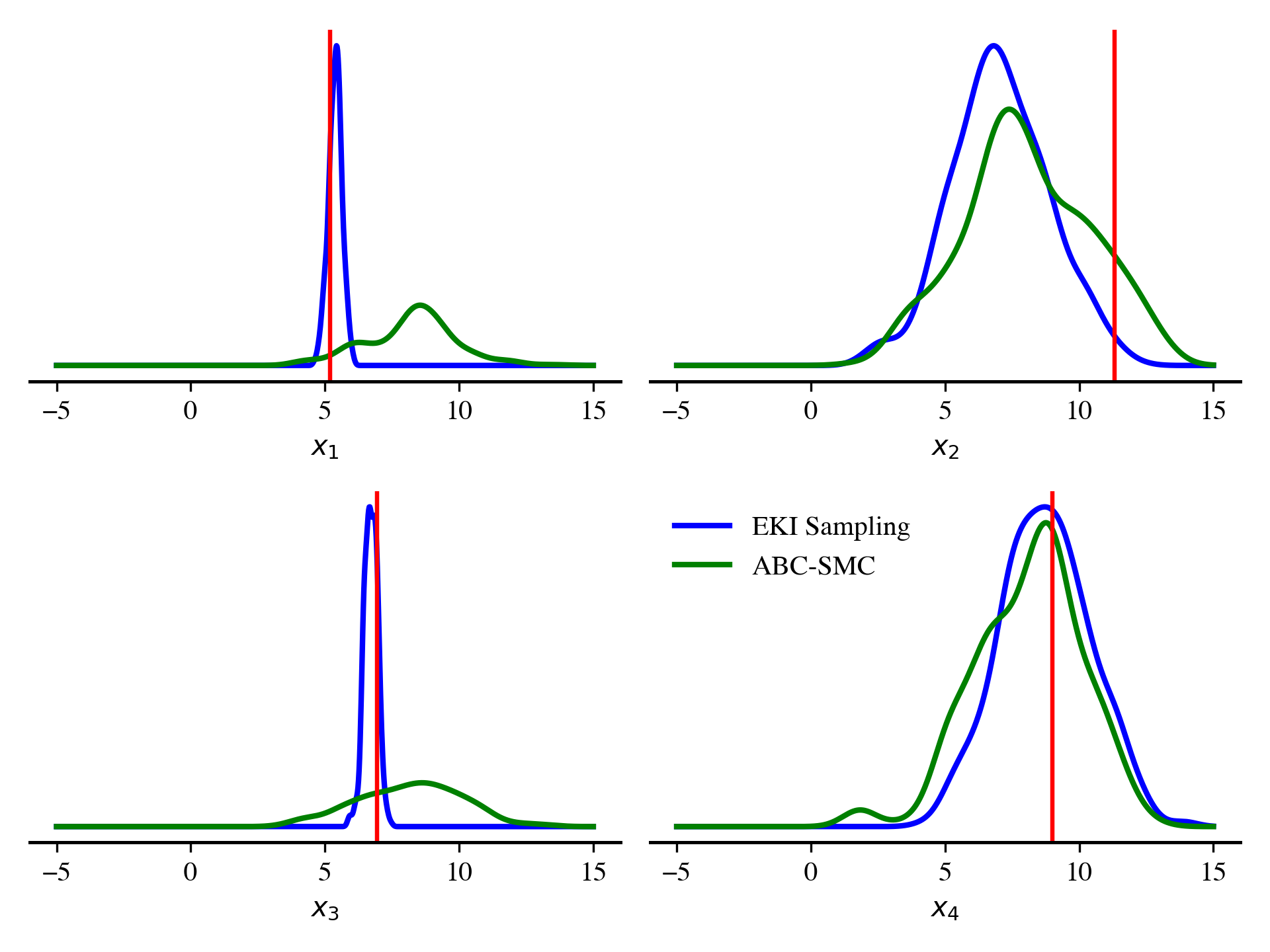}
    \caption{Lorenz 96 marginal densities for EKI sampling and ABC-SMC, $N=500$, truth in red.}
    \label{fig:l96_densities}
    \end{minipage}%
    \begin{minipage}{.48\textwidth}
    \captionsetup{width=.8\linewidth}
    \centering
    \includegraphics[width=0.8\textwidth]{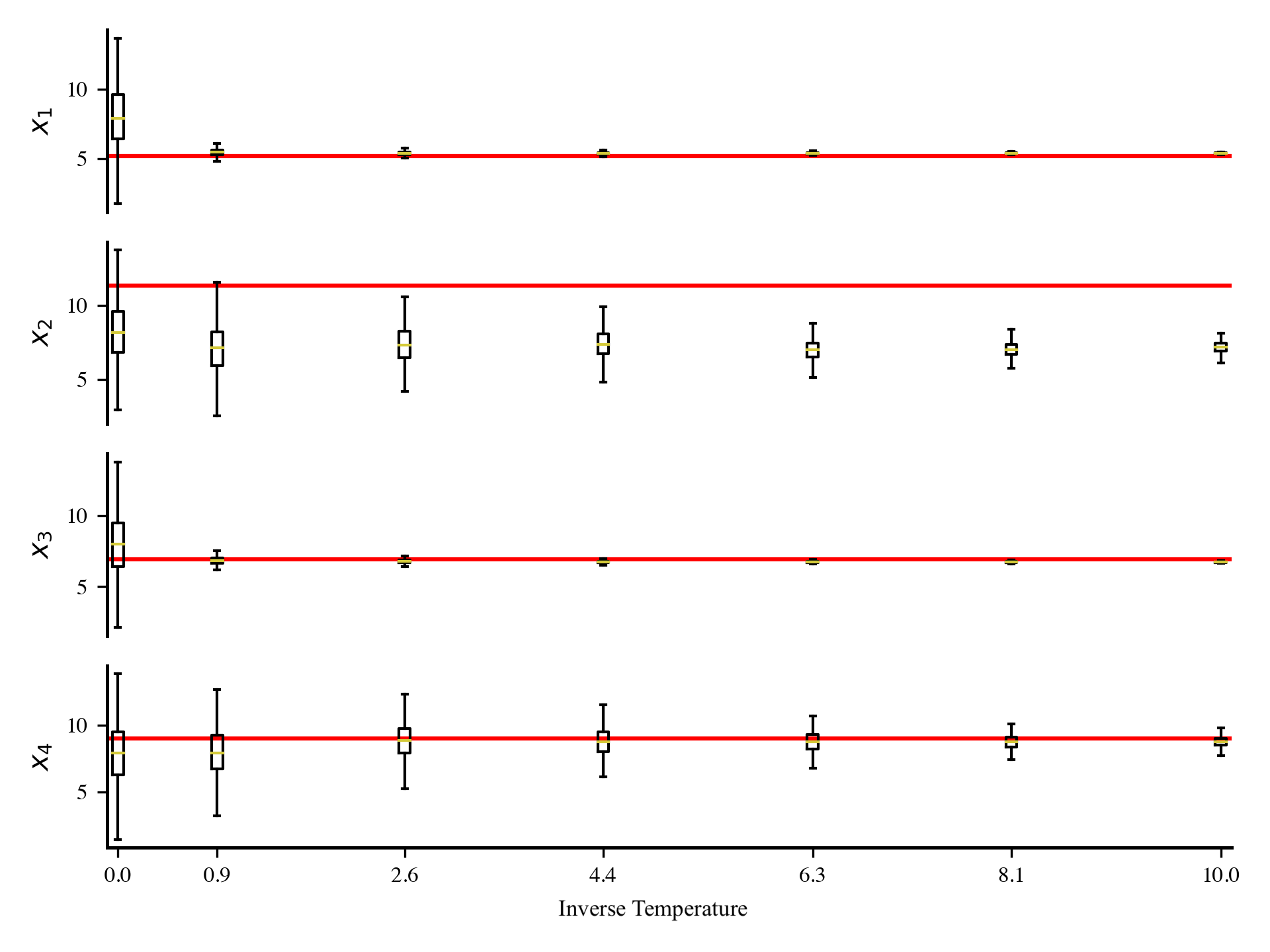}
    \caption{Lorenz 96 coonvergence of EKI for optimisation, $N=500$, truth in red. Boxes display range and quartiles of ensemble.}
    \label{fig:l96_eki_bps}
    \end{minipage}
\end{figure}

We now consider the task of inferring the initial conditions of a noisy version of the Lorenz 96 model \citep{Lorenz95}. The Lorenz 96 model is a simplified representation of oceanic flows that is commonly used as a testbed for high-dimensional data assimilation techniques.
\par
The Lorenz 96 dynamics (with added stochasticity) are defined by the SDE
\begin{equation*}
    dx_t[m] = (-x_t[m-2] \, x_t[m-1] + x_t[m-1] \, x_t[m+1] - x_t[m] + F)dt + dW_t,
\end{equation*}
for $m = 1, \dots, d_x$ with cyclic coordinates $x_t[0] = x_t[d_x]$, $x_t[-1] = x_t[d_x-1]$ and $x_t[d_x + 1] = x_t[1]$. Here the notation $x[i]$ simply accesses the $i$th coordinate of the vector $x$. We adopt the common high-dimensional setup of $d_x=40$ and $F=8$, which is known to produce challenging, chaotic dynamics.
\par

\begin{wrapfigure}{r}{0.5\textwidth}
    \centering
    \includegraphics[width=0.5\textwidth]{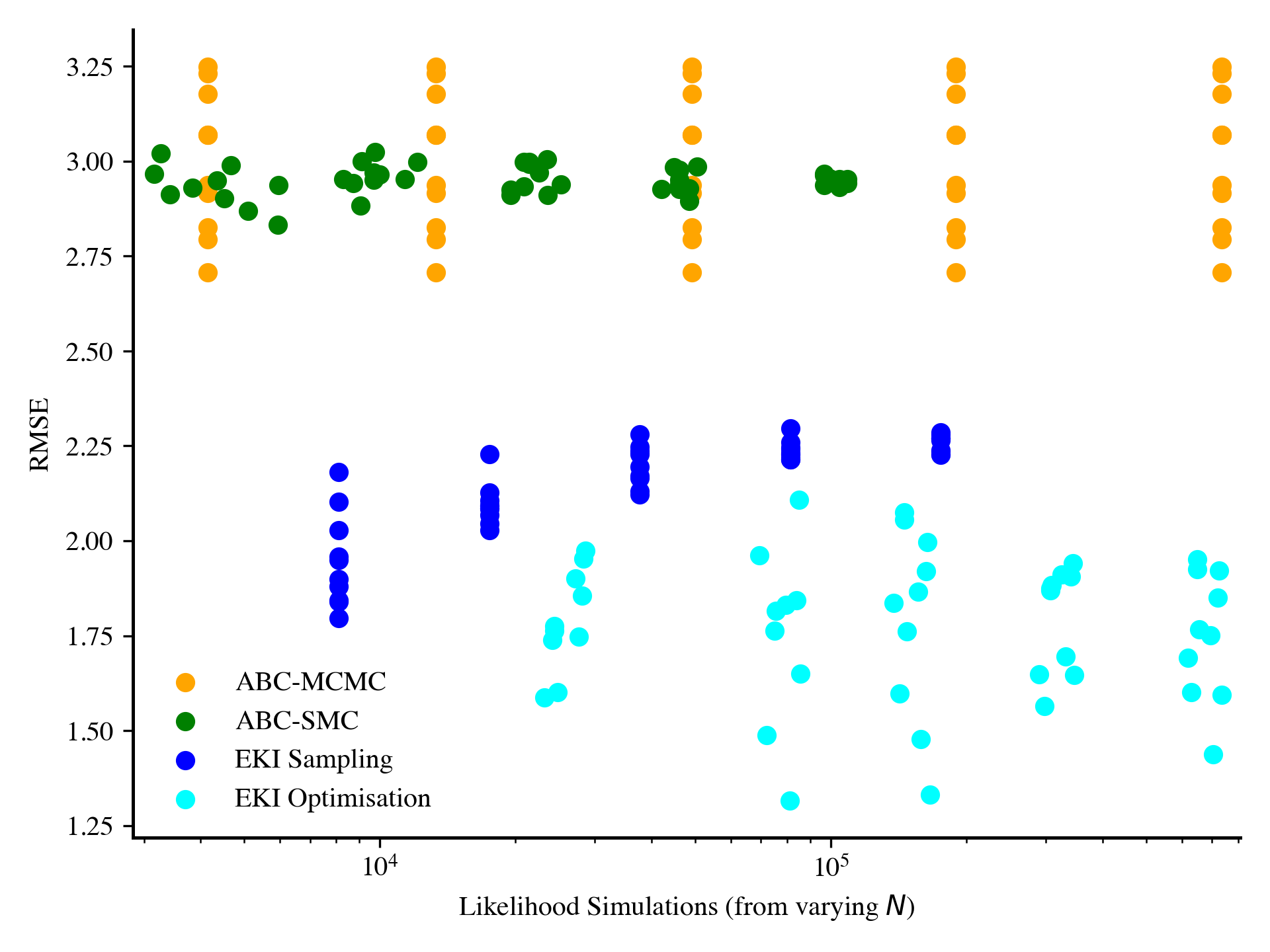}
    \caption{Root mean squared error for number of likelihood simulation induced by varying $N$ (from 200 to 5000), on Lorenz 96 example. Repeated over 10 randomly generated sets of observations.}
    \label{fig:l96_rmse}
\end{wrapfigure}

We define our inference goal as obtaining the initial conditions $x_0 \in \R^{d_x}$ given observations of every other dimension, perturbed by Gaussian noise with variance 0.1, at times $t=1,2,3,4,5$ - resulting in $d_y=100$ observations. We set the prior to $p(x_0) = \gauss(x_0 \mid F, 5\id_{40})$ and simulate from the likelihood with an Euler-Maruyama scheme (stepsize 0.001). The stochasticity in the Lorenz 96 dynamics provides a more realistic influence of uncertainty but the iterative addition of noise in the Euler-Maruyama scheme makes evaluating the likelihood density intractable.
\par
In our experiments, underlying true values for the initial conditions are sampled from the prior and then observations generated using the same Euler-Maruyama scheme as above.


Recall that odd dimensions are directly observed whereas even dimensions are unobserved. We see in \Cref{fig:l96_densities} that EKI for sampling converges very closely around the truth in observed dimensions and is understandably less certain about unobserved dimensions. In contrast ABC-SMC, performs similarly for all dimensions and is likely struggling with the dimensionality of the problem.

When we push the EKI into high inverse temperatures in \Cref{fig:l96_eki_bps} we first see that the particles struggle to collapse in unobserved dimensions. This is an indication that the given observations are insufficient to provide a confident point estimate in those unobserved dimensions. We also notice that the particles converge in the second dimension away from the true underlying value of the parameter - this may be an indication that the true maximum likelihood is not necessarily guaranteed to be close to the truth (for all dimensions) under this model setup.

In \Cref{fig:l96_rmse}, we also observe the phenomenon of decreasing performance in EKI for sampling as $N$ increases for the L96 example - although less so. However, again we see that EKI consistently outperforms ABC, although this is perhaps not surprising in a high-dimensional example given the regular use of ensemble Kalman techniques in similar settings. In this case, the EKI for optimisation utilised significantly more likelihood simulations as it requires more iterations to converge - it also suffered high variance results whereas the performance of the EKI for sampling was more stable.

\section{Conclusion}\label{sec:conc}

We have extended the work of ensemble Kalman inversion \citep{Iglesias2013, Iglesias2018} to problems with general likelihoods as opposed to the common restriction of additive Gaussian noise. In doing so, we remain faithful to the ensemble Kalman paradigm, i.e. our generalisation remains asymptotically unbiased in the linear Gaussian case. We described how to apply the technique for both optimisation and for sampling or uncertainty quantification. We have demonstrated both speed and accuracy of the novel ensemble Kalman inversion algorithm in a difficult benchmark problem as well as a high dimensional spatial example. The computational cost of the ensemble Kalman inversion is $O(Ld_x^3 + LNd_x^2)$ and only requires $LN$ likelihood simulations which is the typical computational bottleneck for problems within approximate Bayesian computation.
\par
We observe the curious phenomenon that increasing the number of particles fails to increase accuracy when applying ensemble Kalman inversion for sampling - at least in terms of mean square error. An outstanding question is how to correct for this. This phenomenon is not entirely new but is not well understood, it would be intriguing to investigate whether it could potentially be mitigated by covariance regularisation \citep{Houtekamer2001} or moment-matching ideas \citep{Lei2011}.
\par
A natural extension of the stochastic ensemble Kalman inversion algorithm presented in this chapter would be the conversion to the square-root ensemble Kalman variants \citep{Bishop2001, Anderson2001} that instead deterministically move particles in a way that remains asymptotically unbiased for fully linear Gaussian problems. By removing a layer of stochasticity we hope to increase the numerical stability of the inversion algorithm.
\par
Outside of linear Gaussian problems, ensemble Kalman inversion is asymptotically biased. It would interesting to investigate embedding an ensemble Kalman kernel within an ABC-SMC sampler. This way, the ensemble Kalman inversion would inherit the theory of approximate Bayesian computation and become asymptotically unbiased for $\nu_\kappa$. However, it is not clear how to choose the backward kernel to induce stable importance weights.
\par
A further application of the presented ensemble Kalman inversion would be to investigate its use within state-space models. It may be that utilising an iterative tempered approach to the update step within an ensemble Kalman filter may improve sample quality. Additionally, we could adapt ensemble Kalman inversion to be used in the online setting of state-space models with intractable observation densities \citep{Jasra2012}.




\section*{Acknowledgements}
This work was supported by the UK Engineering and Physical Sciences Research Council (EPSRC) grant EP/M508007/1 for the Department of Engineering, University of Cambridge.


\singlespacing
\bibliographystyle{elsarticle-harv-nourl}
\bibliography{references}

\end{document}